# Software Effort Estimation from Use Case Diagrams Using Nonlinear Regression Analysis

Ali Bou Nassif
Department of Computer Engineering
University of Sharjah
Sharjah, UAE
anassif@sharjah.ac.ae

Manar AbuTalib
Department of Computer Science
University of Sharjah
Sharjah, UAE
mtalib@sharjah.ac.ae

Luiz Fernando Capretz
Department of Electrical and Computer Engineering
University of Western Ontario
London, ON, Canada
lcapretz@uwo.ca

*Abstract*—Software effort estimation in the early stages of the software life cycle is one of the most essential and daunting tasks for project managers. In this research, a new model based on non-linear regression analysis is proposed to predict software effort from use case diagrams. It is concluded that, where software size is classified from small to very large, one linear or non-linear equation for effort estimation cannot be applied. Our model with three different non-linear regression equations can incorporate the different ranges in software size.

## I. INTRODUCTION AND MOTIVATION

Software cost estimation has been notoriously inaccurate in the last few decades [1]. According to the International Society of Parametric Analysis (ISPA) [2] and the Standish Group International [1], the main reasons behind project failures are:

- Lack of estimation of the staff's skills and levels
- Lack of understanding the requirements
- Improper software size estimation
- Optimism in software estimation

In a nutshell, many software projects fail because of the inaccuracy of software estimation and the misunderstanding or incompleteness of the requirements [3], [4]. The Unified Modeling Language (UML) has been used as a standard to represent the object oriented design. For this reason, software estimators have been recently using UML models to estimate software [5]–[7]. The Use Case Points (UCP) is a prominent method which depends on the use case diagram for software effort estimation [8]. The use case diagram is composed of several use cases which describe the functional requirements of a system. Each use case is represented by a use case scenario which is composed of the Main Success Scenario and Extensions [9]. The main success scenario is the main part of the use case scenario, as it describes the interaction between an actor and a use case in an ideal situation. The Extensions part describes the failure conditions. In the UCP, software size is measured based on the number and the complexity (Simple, Average or Complex) of the use cases as well as the actors in the use case diagram. The effort is estimated by multiplying the software size by a value of twenty. The UCP method has been evaluated in the industry with good results [10], [11]. In general software effort is directly proportional to software size but the relationship is nonlinear. In this paper, we introduce new approaches to predict software size, as well as software effort based on the use case diagrams.

The remainder of the paper is organized as follows: Sections 1.A and 1.B present an overview of evaluation criteria used in this paper and related work. Section 2 proposes the methodology used in this research. Section 3 demonstrates an evaluation of the proposed models. Finally, Section 4 concludes the paper.

### A. Evaluation Criteria

Several methods exist to compare cost estimation models. Each method has its advantages and disadvantages. In our work, four different evaluation methods have been used. These methods include the Mean of the Magnitude of Relative Error (MMRE), the Mean of Magnitude of Error Relative to the Estimate (MMER) the Prediction Level (PRED) and the Mean Error at 95% Confidence Interval (CI).

### B. Related Work

Some work has been done to enhance the effort estimation of the use case point model. Other work was done to build regression models to predict software estimation, and some research used soft computing techniques such as fuzzy logic and neural network models to estimate software effort.

Karner [5] proposed the Use Case Point (UCP) model in 1993. The UCP model is used to predict software effort from the use case diagram based on the number and weight of use cases and actors. The Adjusted Use Case Point (UCP) is calculated by multiplying the Un-adjusted Use Case Point (UUCP) by a variable (usually between 0.7 and 1.3) which represents some Complexity and Environmental factors. Complexity and Environmental factors are listed in Tables 3 and 4 in reference [8]. Finally, software effort is calculated as follows:

$$Effort(Person-Hours) = Size(UCP) * 20. \quad (1)$$

Nassif et al. [12]–[15] enhanced the UCP estimation method by adjusting the weights of the use cases.

Machine Learning models [16]–[21] were also used to improve the accuracy of software size estimation.

This paper is part of the PhD thesis [22]. The main contributions of this paper are as follows:

- We propose a new method to estimate software size from use case diagrams
- We propose a new method to estimate software effort from UCP based on nonlinear regression.

## II. METHODOLOGY

### A. Software Size

There are three main shortcomings when estimating software size using the original UCP model. First, the UCP only uses three categories when classifying use cases (simple, average and complex). A use case is complex when the number of transactions is more than 7. This means a use case of 8 transactions has the same weight of a use case having 20 transactions. Secondly, The UCP assigns the same weight to the Main Success Scenario and the Extensions when they have the same number of transactions. Thirdly, the UCP does not take into consideration the Extend and Include use cases when counting the number of transactions. In this paper, a new approach to predict size estimation is proposed based on the following rules:

- Consider all types of use cases in the use case diagram.
- In the use case scenario of each use case, count the number of transactions in the Main Success Scenario. This is noted by $T_S$.
- In the use case scenario of each use case, count the number of transactions in the Extensions part. This is noted by $T_E$.
- The total number of transactions of the use case is calculated as $T_S + T_E/2$.
- Assign a weight for each use case based on the rules proposed in Table I.
- The total size of the project is conducted by adding the complexity weight of each use case. In other words,

$$Size = \sum_{i=1}^{6} n_i * w_i. \quad (2)$$

Where n is the number of use cases of variety *i* and w is its corresponding weight.

TABLE I.   USE CASE COMPLEXITY

| Complexity | Number of | Complexity |
|---|---|---|
| VL (Very low) | [1,4] | 5 |
| LO (Low) | ]4,8] | 10 |
| NM (Normal) | ]8,12] | 15 |
| HI (High) | ]12 to 16] | 20 |
| VH (Very High) | ]16 to 20] | 25 |
| XH (Extra | > 20 | 30 |

### B. Project Complexity

The complexity of the project is an important factor in software effort prediction. Complex projects require more effort to develop than simple projects that have the same size. The general equation of software effort can be represented as [13]:

$$Effort = \frac{Complexity}{Productivity} \times Size. \quad (3)$$

In our research, we identify the project complexity based on five levels (from Level1 to Level5) as mentioned in [18]:

- "Level1: The complexity of a project is classified as Level1 if the project team is familiar with this type of project and the team has developed similar projects in the past. The number and type of interfaces are simple. The project will be installed in normal conditions where high security or safety factors are not required. Moreover, Level1 projects are those of which around 20% of their design or implementation parts are reusable (came from old similar projects). The weight of the Level1 complexity is 0.7.
- Level2: This is similar to level1 category with a difference that only about 10% of these projects are reusable. The weight of the Level2 complexity is 0.85.
- Level3: This is the normal complexity level where projects are not said to be simple, nor complex. In this level, the technology, interface, installation conditions are normal. Furthermore, no parts of the projects had been previously designed or implemented. The weight of the Level3 complexity is 1.
- Level4: In this level, the project is required to be installed on a complicated topology/architecture such as distributed systems. Moreover, in this level, the number of variables and interface is large. The weight of the Level4 complexity is 1.15.
- Level5: This is similar to Level4 but with additional constraints such as a special type of security or high safety factors. The weight of the Level5 complexity is 1.3."

The weights proposed for each complexity level were based on the thorough analysis conducted on the 212 projects to see how each level would influence the effort estimation. Moreover, some project managers who were involved in developing these projects were consulted to assist us in assigning these weights. Based on this classification, we can say that developing Level5 projects require 30% more effort than Level3 projects.

### C. Productivity

Productivity is inversely proportional to effort. The higher the productivity of a team is, the less effort required to develop a project. We propose five factors to determine the team productivity. Each factor is rated from "1" which represents "very low" to "5" which represents "very high". Factors with average classifications are rated as "3". These factors and their corresponding weights are:

- Team experience regarding the problem domain. Weight is 2.
- Team motivation. Weight is 1.
- Programming language type and experience. Weight is 2.
- Object oriented experience (UML). Weight is 2.
- Analytical skills. Weight is 1.

Regarding the first factor, if the project team is acquainted with the problem domain of the project, the effort required to develop the project will be less than the one if the team is inexperienced with the problem domain. Another important productivity factor is the team experience and the type of the programming language used to implement the project. In general, programmers who are expert in a certain language are those who have at least 5 years of experience. Moreover, the productivity would be higher when using $4^{th}$ generation languages (4GL) such as Visual Basic and Matlab rather than using 3GL such as C++. The team experience in the object oriented concept is very important because the team is either drawing UML diagrams or implementing UML diagrams. This research is based on prediction software estimation from UML use case diagrams. The final factor which contributes to the productivity is the analytical skills of the team..

The second step after assigning a rate (from 1 to 5) to each of the above productivity factors, is to determine the value of the productivity. The productivity factor is calculated in two steps. First, calculate productivity_sum as follows:

$$\mathrm{Productivity\_Sum} = \sum_{i=1}^{5} F_i * W_i. \quad (4)$$

Where $F$ is the productivity factor of variety $i$ and $W$ is its corresponding weight. Based on the rules introduced above, the minimum value of Productivity_Sum is when the rate of all factors is "1". Similarly, the maximum value would be when the rate of all factors is "5". This means that Productivity_Sum falls between 8 and 40. If all productivity factors are average (rate=3), then Productivity_Sum is 24. The second step is to find the final Productivity value which is based on the value of Productivity_Sum as shown in Table II.

TABLE II. PRODUCTIVITY VALUE

| Productivity_Sum | Productivity |
|---|---|
| Less than or equal 14 | 0.7 |
| Between 15 and 20 | 0.85 |
| Between 21 and 27 | 1 |
| Between 28 and 34 | 1.15 |
| Greater than or equal 35 | 1.3 |

The original UCP model assumes that the relationship between software effort and size is linear. As discussed in the Introduction, when the software size increases, software effort will increase but with non-linear relation. To support our hypothesis and to discover the type of this relationship (Effort – Size), among the 212 data projects that we have, 65 projects of software effort ranged between 122 person-hours and 129,353 person-hours were selected that have similar Complexity, Productivity and Requirements Stability. Figure 1 depicts the actual size and effort of these 65 projects as well as the original UCP Estimation. Figure 1 shows that the UCP method can be applied with acceptable error on small projects (size less than 250 UCP which is equivalent to 5,000 person-hours). Based on Figure 1, the UCP model cannot be applied on projects of effort more than 10,000 person-hours. Among the 212 data projects that we have, there are 56 projects (26%) that have effort more than 10,000 person- hours. This means that projects of greater than 10,000 person-hours cannot be ignored. The plot of the actual data projects in Figure 1 show that the relationship between software effort and size in non-linear. However, many non-linear functions exist and it is not simple to just predict one. Based on the nature of this non-linear relationship, we used four different non-linear equations to see which equation can best fit the actual data. These equations include a second degree polynomial function and three exponential functions as shown in Table III, where the variable "x" corresponds to software size, the variable "y" corresponds to software effort and "a", "b", "c" and "d" are constants.

TABLE III. NON-LINEAR EQUATIONS

| Polynomial | Exponential 1 | Exponential 2 | Exponential 3 |
|---|---|---|---|
| $y = a * x^2 + b * x + c$ | $y = a + b * \exp(c * x)$ | $y = a * \exp(b * x) + c * \exp(d * x)$ | $y = \exp(a + b * x)$ |

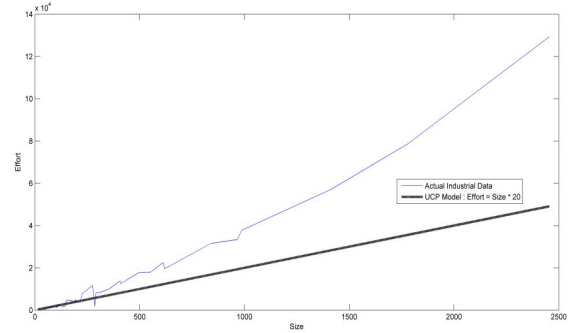

Fig. 1. Comparison Between UCP Model and Actual Data

In each non-linear equation type (Table III), several experiments were conducted using the whole dataset to calculate the values of the constants "a", "b", "c" and "d". In each experiment, the value of the coefficient of determination $R^2$ and the Root Mean Square (RMS) were measured. $R^2$ is the percentage of variation in Effort explained by the variable Size. An acceptable value of $R^2$ is $\geq 0.5$ [14].

The whole project dataset that is used to build the non-linear regression models, (65 projects) was divided into three different ranges based on the software size. The first range is called Small, which includes 26 projects out of the 65 projects of software size less than 100 UCP (less than 2,000 person-hours). The second range is the Medium range that contains 21 projects of size ranged between 100 and 300 UCP (between 2,000 and 8,500 person-hours) and the third range is the Large one which contains 18 projects of size greater than 300 UCP (effort between 8,500 and 129,353 person-hours). Several experiments were performed to learn which of the four non-linear equations (Table III) can best fit each range. Experiments show that based on the fitting graphs, values of $R^2$ and RMS, the Polynomial model can best fit the small dataset. However, the Exponential 3 and Exponential 2 models can best fit the Medium and the Large ranges, respectively.

## III. MODEL EVALUATION

We used 146 projects for testing our model. The evaluation of our model was conducted through four different experiments. First, the whole dataset (146 projects) was used. Then, we divided the whole dataset into three ranges which include 59 small projects of size less than 100 UCP, 48 medium-sized projects of size between 100 and 300 UCP and 39 large projects of size greater than 300 UCP. In each of the four experiments, our model was evaluated against other models that predict software estimation from the use case diagrams such as the UCP model (Table IV). The evaluation criteria used for testing are MMER, MMER, and Mean Error with CI at 95%, as well as PRED (25), PRED (50), PRED (75) and PRED (100).

TABLE IV. NON-LINEAR EQUATIONS

| Criteria | UCP | | | | Proposed Nonlinear Regression | | | |
|---|---|---|---|---|---|---|---|---|
| | All | Sm | Md | Lg | All | Sm | Md | Lg |
| MMRE | 0.60 | 0.58 | 0.62 | 0.60 | 0.46 | 0.47 | 0.44 | 0.47 |
| MMER | 2.02 | 2.35 | 2.10 | 1.41 | 0.77 | 1.13 | 0.70 | 0.31 |
| PRED (25) | 9.58 | 1.69 | 14.5 | 15.3 | 21.2 | 8.47 | 14.58 | 48.71 |
| PRED (50) | 19.1 | 11.8 | 22.9 | 28.2 | 45.8 | 22.0 | 50 | 76.9 |
| PRED (75) | 33.5 | 32.2 | 29.1 | 41.0 | 65.7 | 40.6 | 70.8 | 97.4 |
| PRED (100) | 43.8 | 40.6 | 37.5 | 56.4 | 73.2 | 55.9 | 75 | 97.4 |
| CI(95%) | 4,48 ± 1,30 | 1,70 ± 725 | 4,24 ± 1,27 | 8,98 ± 4,18 | 968 ± 936 | 1,37 ± 614 | 2,44 ± 968 | -1,4 ± 3 |

## IV. CONCLUSIONS

It is clear from Table IV that the proposed model outperforms the original UCP model based on 7 criteria. Lower values of MMRE and MMER indicate better accuracy. On the contrary, higher values of PRED(x) indicate better accuracy.